\documentclass[AMA,Times2COL]{WileyNJDv5}
\usepackage{subfig}
\usepackage{comment}
\articletype{Reserach Article}%

\received{17 February 2025}
\revised{Date Month Year}
\accepted{Date Month Year}
\doiheadtext{\url{https://doi.org/10.1002/jor.26111}}
\journal{Journal of Orthopaedic Research}

\volume{n/a}
\copyyear{2025}
\startpage{1}

\begin{document}

\title{Improved Accuracy in Pelvic Tumor Resections Using a Real-Time Vision-Guided Surgical System}

\author[1]{Vahid Danesh}
\author[2]{Paul Arauz}
\author[1]{Maede Boroji}
\author[2]{Andrew Zhu}
\author[1]{Mia Cottone}
\author[3]{Elaine Gould}
\author[2]{Fazel A. Khan}
\author[1]{Imin Kao}

\authormark{DANESH \textsc{et al.}}
\titlemark{Improved Accuracy in Pelvic Tumor Resections Using a Real-Time, Vision-Guided Surgical System}

\address[1]{\orgdiv{Department of Mechanical Engineering}, \orgname{Stony Brook University}, \orgaddress{\state{NY}, \country{USA}}}
\address[2]{\orgdiv{Department of Orthopedics, Renaissance School of Medicine}, \orgname{Stony Brook University}, \orgaddress{\state{NY}, \country{USA}}}
\address[3]{\orgdiv{Department of Radiology, Renaissance School of Medicine}, \orgname{Stony Brook University}, \orgaddress{\state{NY}, \country{USA}}}

\corres{Corresponding author Imin Kao, Department of Mechanical Engineering, Stony Brook University, Stony Brook, NY, USA. \email{imin.kao@stonybrook.edu}}



\abstract[Abstract]{Pelvic bone tumor resections remain significantly challenging due to complex three-dimensional anatomy and limited surgical visualization. Current navigation systems and patient-specific instruments, while accurate, present limitations including high costs, radiation exposure, workflow disruption, long production time, and lack of reusability. This study evaluates a real-time vision-guided surgical system combined with modular jigs to improve accuracy in pelvic bone tumor resections. A vision-guided surgical system combined with modular cutting jigs and real-time optical tracking was developed and validated. Five female pelvis sawbones were used, with each hemipelvis randomly assigned to either the vision-guided and modular jig system or traditional freehand method. A total of twenty resection planes were analyzed for each method. Accuracy was assessed by measuring distance and angular deviations from the planned resection planes.
The vision-guided and modular jig system significantly improved resection accuracy compared to the freehand method, reducing the mean distance deviation from 2.07 $\pm$ 1.71 mm to 1.01 $\pm$ 0.78 mm (\emph{P} = .0193). In particular, all specimens resected using the vision-guided system exhibited errors of less than 3 mm. Angular deviations also showed significant improvements with roll angle deviation reduced from 15.36 $\pm$ 17.57$^\circ$ to 4.21 $\pm$ 3.46$^\circ$ (\emph{P}=.0275), and pitch angle deviation decreased from 6.17 $\pm$ 4.58$^\circ$ to 1.84 $\pm$ 1.48$^\circ$ (\emph{P}<.001). The proposed vision-guided and modular jig system significantly improves the accuracy of pelvic bone tumor resections while maintaining workflow efficiency. This cost-effective solution provides real-time guidance without the need for referencing external monitors, potentially improving surgical outcomes in complex pelvic bone tumor cases.}


\jnlcitation{\cname{%
\author{Danesh V.},
\author{Arauz P.},
\author{Boroji M.},
\author{Zhu A.},
\author{Cottone M.},
\author{Gould E.},
\author{Khan F. A.}, and
\author{Kao I.}}.
\ctitle{Improved Accuracy in Pelvic Tumor Resections Using a Real-Time, Vision-Guided Surgical System} \cjournal{\it Journal of Orthopaedic Research®. \url{https://doi.org/10.1002/jor.26111}} \cvol{2025; 1-8}.}

\maketitle



\section{Introduction}
\label{sec:introduction}
Pelvic bone tumor surgeries are highly complex due to the intricate morphology of the pelvic bones and the critical nature of surrounding tissues. Ensuring safe surgical margins is essential to reduce local recurrence rates, which can be as high as 70\% for marginal resections and 90\% for intralesional resections~\cite{ozaki2003osteosarcoma, fuchs2009osteosarcoma}.

With current advanced imaging modalities, orthopedic surgeons can precisely define the boundaries of primary bone tumors prior to surgical resection~\cite{hoffer2000, saifuddin2002, khan2013}. Although surgeons can develop an ideal and precise resection plan preoperatively to adhere to oncologic principles~\cite{springfield1984,springfield1988} and preserve healthy tissue, conventional tools and techniques, such as freehand procedures, often can not provide reliable and consistent execution of the preoperative plan during surgery~\cite{khan2013,cartiaux2008surgical,cartiaux2010computer}, potentially leading to serious consequences~\cite{khan2013}. For example, a surgeon might inadvertently cut into the tumor during the procedure. In cases like osteogenic sarcoma, such imprecision can lead to higher rates of local recurrence and increased mortality~\cite{bacci2005,nathan2006}. Alternatively, the surgeon might choose to resect significantly more healthy tissue to avoid tumor disruption. Although this approach adheres to oncologic principles, the removal of excessive normal tissue can significantly impact functional outcomes and recovery~\cite{avedian2010}.



Computer-assisted navigation has been developed for pelvic bone tumor surgeries with the main objective of improving bone cutting accuracy~\cite{cartiaux2013, cartiaux2011, abraham2011, fehlberg2009, garcia2021}. Several navigation systems are available for surgeons to benefit from a real-time navigated process to improve bone tumor cutting accuracy. A previous study on computer-assisted navigation for simulated pelvic bone tumor surgery demonstrated that the bone cutting errors were twice smaller when compared to a solely freehand method~\cite{cartiaux2013}. However, the adoption of surgical navigation systems for pelvic bone tumor resection has been limited by high costs and workflow disruptions. 

Patient Specific Instrument (PSI) method has been developed as an alternative to intraoperative navigation. PSIs have been used for pelvic bone tumor surgery with customized cutting guides to restrict the trajectories of the cutting tool around the tumor~\cite{biscaccianti2022, cartiaux2014, fragnaud2022, jentzsch2016, sallent2017}. Customized cutting guides have been described as a reliable surgical aid, minimizing bone resection and reducing local recurrence compared to freehand tumor resection procedures~\cite{cartiaux2014, fragnaud2022, evrard2019}. Their use may be especially relevant in the pelvis due to the depth of dissection and complex spatial relationships between critical anatomic structures~\cite{fragnaud2022,mavrogenis2012}. However, this technology often encounters challenges with preoperative delays due to the time required for planning and production, high manufacturing costs driven by the customization process and regulatory compliance~\cite{dorling2023cost}, and the difficulty of achieving accurate positioning intraoperatively.

The objective of this experimental study is to evaluate the accuracy of simulated pelvic bone tumor resections by integrating a  validated light-projection navigation system with a modular cutting jig into a clinical-grade vision-guided surgical system~\cite{he2022light-jigs, he2022light3d, he2022novelVPS}. This system uses cost-effective components to integrate intuitively the pelvic bone tumor resections into the surgical workflow, and provide real-time guidance comparable to computer navigation systems~\cite{he2022light3d, he2021registration}. Unlike traditional computer image-guided surgery, this system eliminates the need for surgeons to constantly reference a separate preoperative plan displayed on a monitor, thereby streamlining the surgical process~\cite{he2022light-jigs}. We hypothesize that the vision-guided surgical system will provide superior accuracy in replicating preoperative plans compared to the traditional freehand method, thereby reducing the risk of local recurrence and preserving more healthy tissue.

\section{Material and Methods}
\label{sec:material}

\subsection{Study Design}
Five female pelvis sawbones (ten hemipelvises) were used in this study. Each hemipelvis was randomly assigned to either the traditional freehand surgical method or the vision-guided surgical system to assess and compare their accuracy in performing pelvic resections.

\subsection{Tumor Modeling} 

\begin{figure*}
    \centering
    \includegraphics[width=0.99\linewidth]{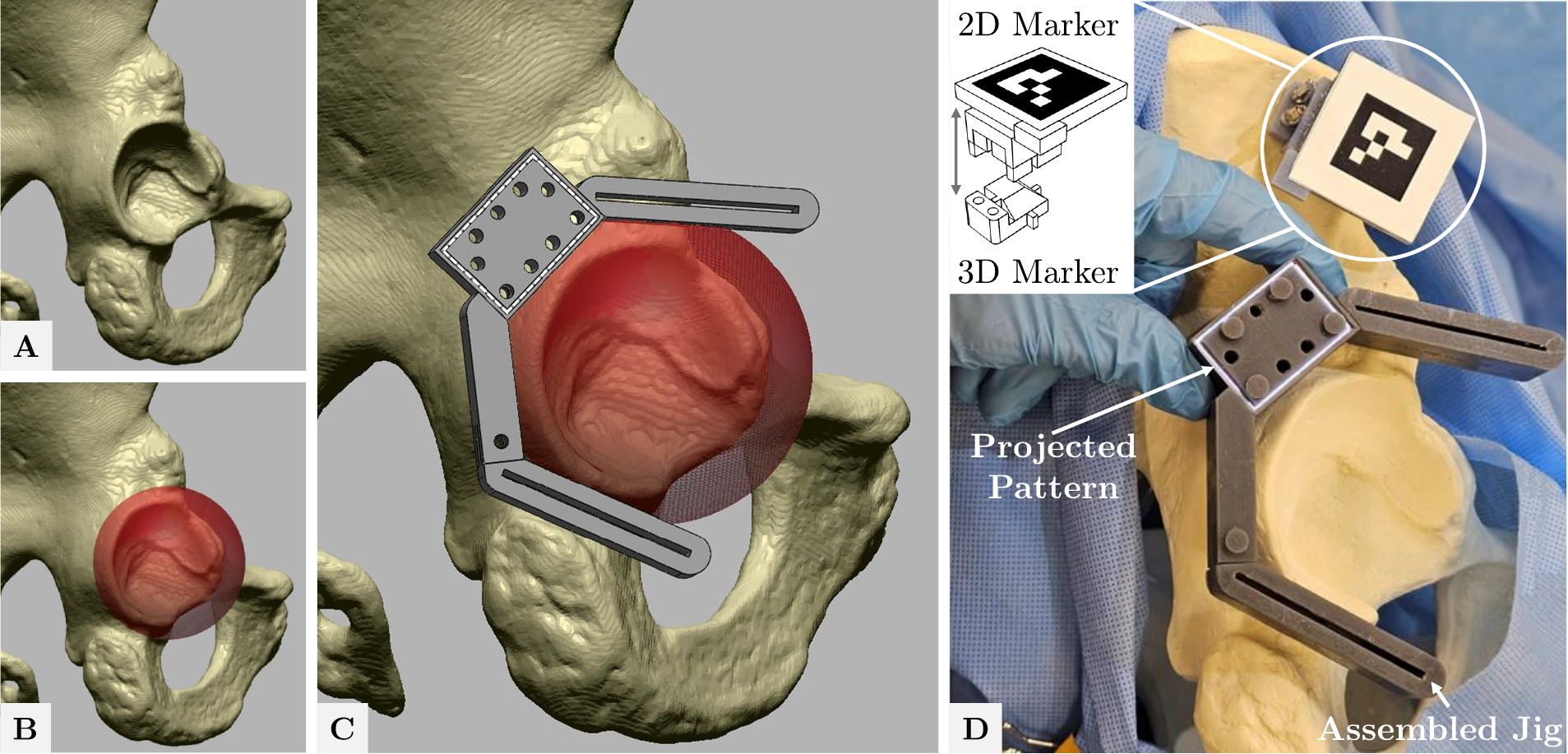}
    \caption{Preoperative and intraoperative planning for type II tumor resection. (A) CT images are converted to a 3D model for preoperative planning.
    (B) A virtual sphere is centered on the hip rotation center to simulate a Type II tumor resection scenario, considering an adequate safety margin.
    (C) Modular jig components are placed on the bone model to encompass the tumor and safety margin. The white dashed lines show the engraved rectangular pattern on the modular jig to align with projected pattern. 
    (D) The assembled jig is aligned with the projected pattern during surgery. The upper left inset shows the “3D-Marker” and “2D-Marker,” which are used for bone registration and real-time tracking, respectively.}
    \label{fig:preoperative}
\end{figure*}

Computed Tomography (CT) images of pelvis sawbones were acquired and converted into Digital Imaging and Communications in Medicine (DICOM) files. These files were then used to generate 3D bone models for further processing, as shown in Figure~\ref{fig:preoperative}A. Virtual Spheres were centered on the hip rotation center to represent tumors in the periacetabular region (Type II pelvic resection scenario, Fig~\ref{fig:preoperative}B). Tumors of the acetabulum are difficult to treat and have a high complication and morbidity rate~\cite{Bickels2001}. Traditional freehand resection approaches for this type of tumor commonly require performing at least three osteotomies: superior pubic ramus, infra-acetabular, and supra-acetabular~\cite{Bickels2001, Malawer2001}.

\subsection{Resection Procedures}

The study includes two independent methods, as follows, to compare the results of resections.

\begin{itemize}
    \item The freehand method: The surgeon reviewed the CT scan and preoperative plan in a 3D modeling format on a computer monitor, with no time limitation. Key measurements relative to bony landmarks were provided. Using standard operating room tools like rulers and protractors, the surgeon marked the osteotomy lines on the bone surface with a marking pen. A power surgical saw (Stryker) with a 1.27~mm thick and 25~mm wide blade was then used to perform the resection.

    \item The vision-guided system: After reviewing preoperative 3D models of the tumor, surrounding tissues, and bone, the surgeon selected the standard snap-fit components from stock. Following the steps depicted in Appendix~\ref{app:modular_jig}, the assembled jig was placed on the 3D bone model to encase the tumor while maintaining the necessary safety margin (Fig~\ref{fig:preoperative}C). During surgery, the final jig can be quickly assembled using standard modular components (Fig~\ref{AppFig:Jig}). After exposing the bone, a small “3D-marker” was affixed to the bone for bone registration. The vision-guided system, comprising a 3D surface scanner, a machine-vision camera, and a light projector, was positioned overhead the surgical table. The 3D scanner registered the bone with the pre-operative CT-scan model by identifying the 3D-marker's position and orientation. A “2D-marker” was then snap-fit onto the 3D-marker to enable real-time tracking of the bone without re-registration during the surgery. The projector system projected the pre-defined pattern onto the surgical area. The surgeon aligned the modular jig with the projected pattern (Fig~\ref{fig:preoperative}D) and, once satisfied, pinned it to the bone with two parallel K-wires. The surgeon then used the built-in slots on the modular jig to guide the resection, as depicted in Appendix~\ref{app:modular_jig}. 
    
\end{itemize}

The experiments were conducted by a resident surgeon (AZ), who was not involved in the development of the vision-guided system. Prior to the experiments, the surgeon was thoroughly introduced to the method and asked to perform resections without bias.

\subsection{Measurements of Accuracy}
\label{sec:measurment}
\begin{figure}[!htbp]
    \centering
    \includegraphics[width=1\linewidth]{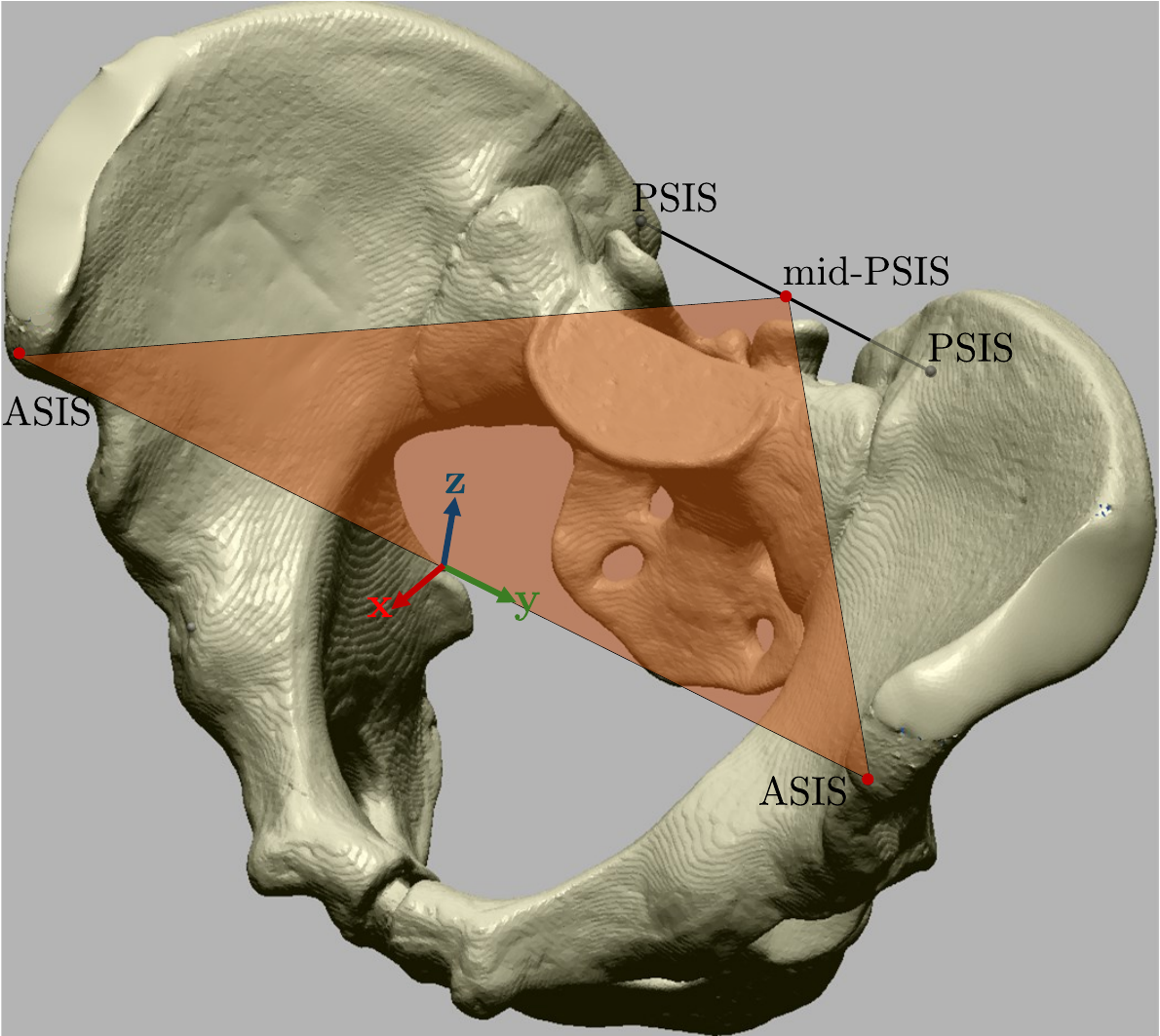}
    \caption{Pelvic Coordinate System. The superior iliac spine plane (SISP), defined by the ASISs and the midpoint of the PSISs, is shown in orange. The ASISs and the midpoint of the PSISs are depicted in red, and the PSISs in black.}
    \label{fig:csys}
\end{figure}

To accurately measure angular deviation, the pelvic coordinate system is defined with the origin at the midpoint of anterior superior iliac spines (ASISs). The positive Y-axis is oriented in the direction connecting the left and right ASISs. The X-axis is defined by the line connecting the midpoint between the ASISs to the midpoint between the posterior superior iliac spines (PSISs), pointing ventrally and orthogonal to the Y-axis. The Z-axis is perpendicular to both the X and Y axes, pointing cranially, as shown in Figure~\ref{fig:csys}.

\begin{enumerate}
\item \textbf{Distance deviation:}
    Distance deviation, in millimeter,  is defined as: 
    \[\textit{Distance Deviation} = \left|M_p - M_r\right|\] 
    where $M_p$ is the margin distance obtained at the closest point between the planned cutting planes and the boundary of the virtual tumor, and $M_r$ is the margin distance at the closest point between the post-operative resected planes and the boundary of the virtual tumor. A smaller distance deviation indicates a more precise adherence to the preoperative plan, ensuring that the resection closely follows the intended surgical margins.
    
\item \textbf{Roll angle:}
    Angular roll deviation, in degrees, is defined as the rotation about the bone's anteroposterior axis (X-axis)  between the planned target planes and resected planes.    

\item \textbf{Pitch angle:}
    Angular pitch deviation, in degrees, is defined as the rotation about the bone's mediolateral axis (Y-axis) between the planned target planes and resected planes. 
    
\item \textbf{Surgical margin error:}
    In clinical practice, the term “positive margin” refers to the presence of tumor cell at the boundary of the resected tissue, indicating incomplete removal. Conversely, a “negative margin” indicates no tumor cell at the resection boundary, implying complete resection. In bone tumor resection, surgeons also establish a “safety margin”, which is a predetermined buffer of healthy tissue removed around the tumor site to compensate for surgical deviations. By evaluating the maximum deviation observed in this study, we can determine the likelihood of maintaining a negative margin during surgery. This error margin guides the surgeon in maintaining an adequate “safety margin” to avoid intralesional resection.
\end{enumerate}

\subsection{Statistical Analysis}
A total of 20 resection planes were analyzed for each method. Statistical comparisons between the vision-guided surgical system and traditional freehand approach focused on geometric precision parameters (distance and angular deviations) relative to the preoperative plan. Mean, standard deviation, and range were calculated for each parameter. The Wilcoxon rank-sum test was used to assess statistical significance between the two approaches, with significance level set at $\alpha = 0.05$. 

\begin{table}%
\caption{Summary of results of experimentally using the traditional freehand method and the vision-guided surgical system\label{table:results}}
\begin{tabular*}{\linewidth}{@{\extracolsep\fill}lrrrr@{}}
\toprule
\textbf{Approach} & \textbf{DD~(mm)} & \textbf{RD~($^\circ$)} & \textbf{PD~($^\circ$)} & \textbf{MD~(mm)} \\ \midrule
\multicolumn{5}{l}{\textbf{Vision-guided system}} \\ 
Mean & 1.01 & 4.21 & 1.84 & 1.92 \\ 
SD   & 0.78 & 3.46 & 1.48 & 0.60 \\ 
Max.  & 2.43 & 11.61 & 6.12 & 2.43 \\ 
Min.  & 0.08 & 0.27 & 0.02 & 0.88 \\ 
\\
\multicolumn{5}{l}{\textbf{Freehand method}} \\ 
Mean & 2.07 & 15.36 & 6.17 & 3.77 \\ 
SD   & 1.71 & 17.57 & 4.58 & 2.37 \\ 
Max.  & 7.43 & 61.04 & 19.34 & 7.43 \\ 
Min.  & 0.05 & 0.08 & 0.22 & 1.46 \\ 
\\
{\em P}-value & \textbf{.0193} & \textbf{.0275} & \textbf{$<$.001} & .0952 \\ 
Sample size & 20 & 20 & 20 & 5 \\ 
\bottomrule
\end{tabular*}
\begin{tablenotes}
Abbreviations: SD, standard deviation; DD, distance deviation; RD, roll angle deviation; PD, pitch angle deviation; MD, maximum deviation: Max., maximum; Min., minimum.

\end{tablenotes}
\end{table}

\begin{table}%
    \centering
    \caption{Likelihood of maintaining a negative (clean) margin during surgery with various thresholds.}
    \renewcommand{\arraystretch}{1.5}
    \begin{tabular*}{\linewidth}{@{\extracolsep\fill}lccc@{}}
    
        \hline
        \textbf{Approach} & \multicolumn{3}{c}{\textbf{Margin Thresholds}} \\
        \cline{2-4}
        & <1 mm & <3 mm & <5 mm \\
        \hline
        Vision-guided surgical system & 60\% & 100\% & 100\% \\
        Freehand method & 30\% & 85\% & 95\% \\
        \hline
    
    \end{tabular*}
    \renewcommand{\arraystretch}{1}
    \label{tab:margin_error}
\end{table}

\begin{figure*}[!htbp]
    \centering
    \includegraphics[width=0.99\linewidth]{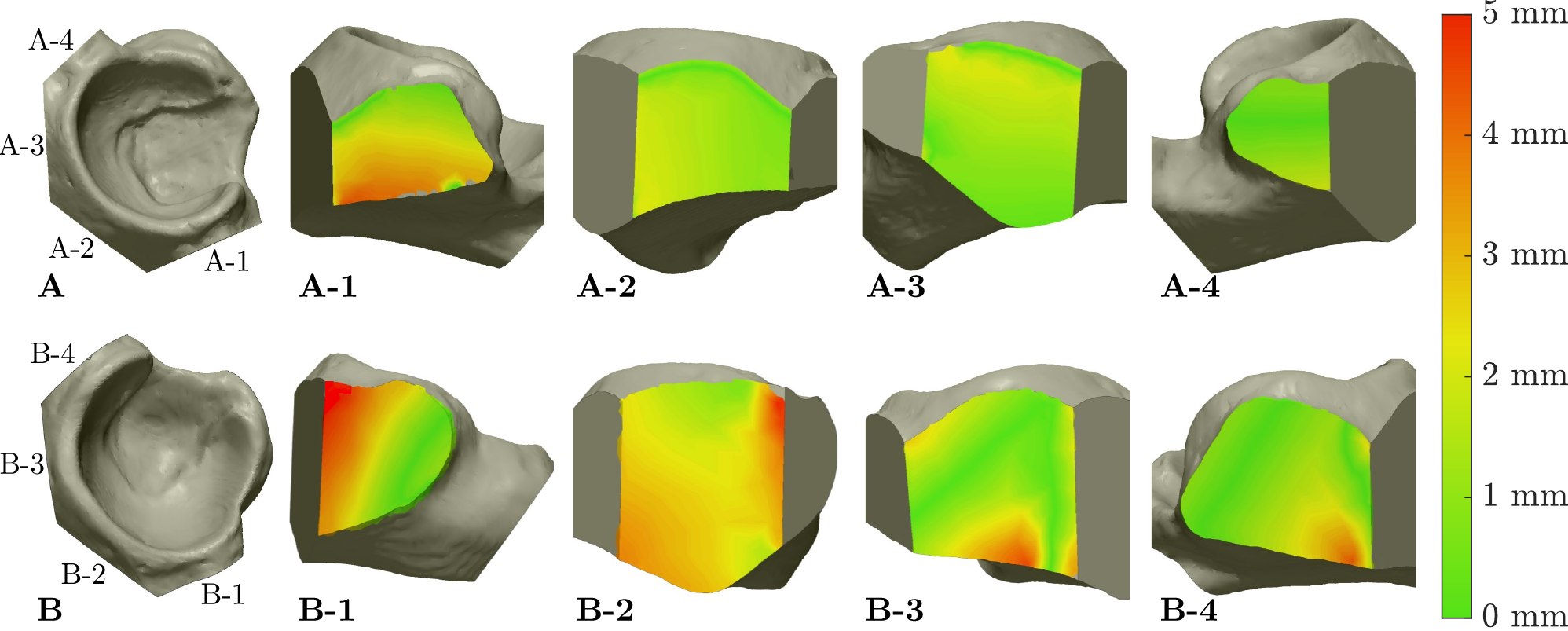}
    \caption{Distance error distribution (mm) between planned and resected planes measured along the normal of each cutting plane for a randomly selected bone comparison, comparing vision-guided surgical system (A) and traditional freehand method (B). A-1 through A-4 demonstrate lower distance deviation and higher resection accuracy compared to B-1 through B-4. Red coloring and inconsistent color distribution across planes in B indicate higher roll and pitch angle deviations compared to the vision-guided system.}
    \label{fig:distanceDeviation}
\end{figure*}

\section{Results}
Compared to the traditional freehand method, the vision-guided surgical system significantly improved the surgeon's ability to replicate the preoperative plan for type II pelvic bone tumor resection across three of the four measures (Table~\ref{table:results}). The mean distance deviation was reduced by 51.2\% from 2.07 $\pm$ 1.71 mm to 1.01 $\pm$ 0.78 mm (\emph{P}=.0193). The mean roll angle deviation decreased by 72.6\% from 15.36$^\circ$ $\pm$ 17.57$^\circ$ to 4.21$^\circ$ $\pm$ 3.46$^\circ$ (\emph{P}=.0275). Similarly, the mean pitch angle deviation was reduced by 70.2\% from 6.17$^\circ$ $\pm$ 4.58$^\circ$ to 1.84$^\circ$ $\pm$ 1.48$^\circ$ (\emph{P}<.001). The maximum deviation indicates no significant improvement in surgical precision. Mean differences from 3.77 $\pm$ 2.37 mm to 1.92 $\pm$ 0.60 mm (\emph{P}=.0952) were observed.

The surgical margin error analysis (Table~\ref{tab:margin_error}) confirms the superior capability of the vision-guided surgical system in maintaining required safety margin, with likelihood of 60\% to maintain a 1mm margin compared to 30\% with the freehand method. Most significantly, while the vision-guided surgical system achieved a 100\% success rate in maintaining a 3mm margin, the freehand method showed an 85\% success rate at 3mm and, critically, 95\% success rate at the 5mm threshold, indicating risk of intralesional resection. 

Figure~\ref{fig:distanceDeviation} illustrates the distribution of distance errors (mm) across each resection plane between preoperative planned and resected planes for a randomly selected bone comparison, highlighting the vision-guided surgical system's lower deviations and higher accuracy compared to the freehand method. The vision-guided surgical system shows consistent color distribution, indicating minimal roll and pitch angle deviations, whereas the freehand method displays marked deviations, as evidenced by red coloring and inconsistent patterns.

\begin{figure}[ht!]
\centering
\includegraphics[width=1 \linewidth]{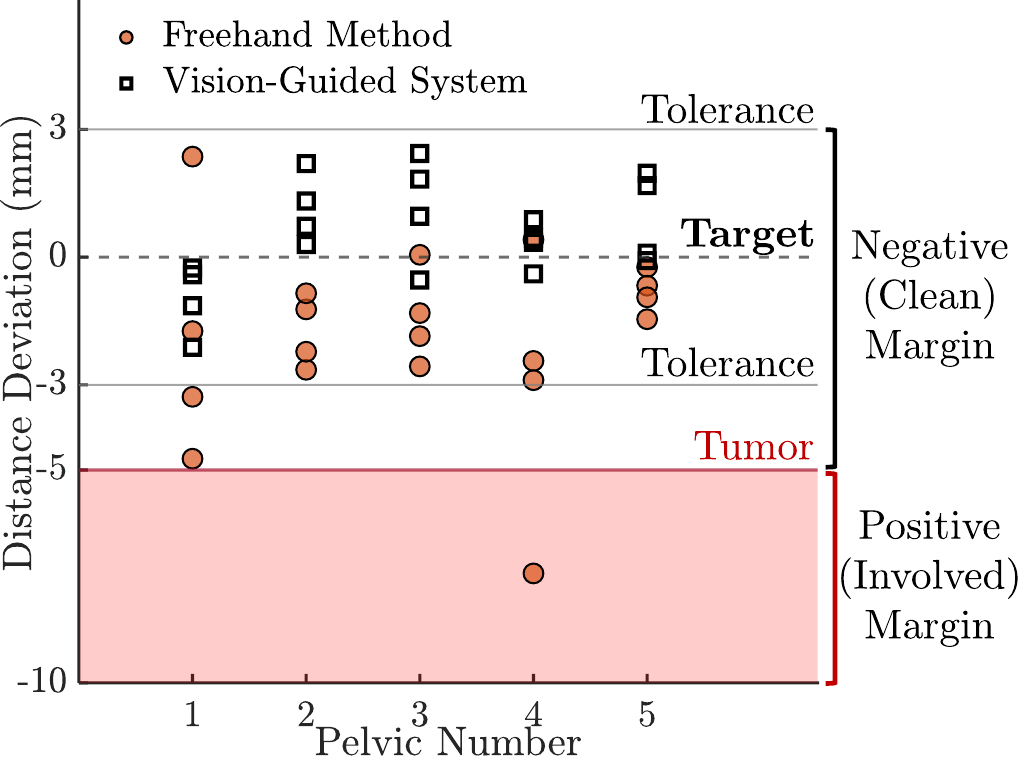}
\caption{Distance deviation (mm) comparison between freehand method, in circles, and vision-guided system, in squares, each marker represents a resection plane The vision-guided system consistently maintains a 3mm tolerance, while the freehand method results in intralesional resections. The red area represents the tumor, with deviations exceeding 5mm classified as positive (involved) resections.}
    \label{fig:distance}
\end{figure}

Figure~\ref{fig:distance} illustrates the actual distance deviations, with negative values indicating resections closer to the simulated tumor. A 3mm threshold was used to assess the capability to maintain the required safety margin during surgery. The vision-guided surgical system consistently maintained this 3mm safety margin across all specimens. In contrast, the freehand method resulted in 15\% (3 out of 20) of resections exceeding the 3mm error threshold, and 5\% (1 out of 20) resulted in intralesional resection.

\section{Discussion}
Accurate reproduction of preoperative plans is crucial in bone sarcoma resections, yet traditional methods often lack enough precision to maintain required safety margin~\cite{cartiaux2010computer, cartiaux2008surgical}. This study highlights the effectiveness of a novel vision-guided surgical system, which significantly outperforms the traditional freehand approach in replicating preoperative plans, as evident by improvements in all measured parameters (Table~\ref{table:results}). Therefore, this study confirms the hypothesis that the vision-guided and modular jig system outperforms conventional freehand procedures. 

Our findings show that the vision-guided system performs comparably to previous studies using computer navigation systems~\cite{cartiaux2013} and 3D-printed patient specific instruments~\cite{fragnaud2022}, with maximum distance deviations ranging from 2.8 mm~\cite{cartiaux2013} to 4 mm~\cite{fragnaud2022}. In addition, the results of this study demonstrate that the performance of the vision-guided modular jig system remains consistent, even when compared to our previous studies on femur bone sarcoma resections~\cite{he2022light-jigs}. Despite the more complex morphology of the pelvic bones, the performance of the vision-guided and modular jig system was unaffected. Furthermore, these results demonstrate that the vision-guided and modular jig system not only reduced all errors but also improved consistency across resections, as evident by the lower standard deviations. The system's capability to maintain accuracy and reduce variability indicates its strong potential to significantly improve surgical outcomes. Our findings suggest that this innovative technology could be effectively implemented in clinical settings, providing both oncological and functional benefits to patients.

While computer navigation has improved surgical accuracy, they come with high costs and complex setup, often involving radiation exposure. Our vision-guided and modular jig system offers a cost-effective and compact alternative, utilizing a  standard light projector, a video camera, and structured-light 3D scanner to achieve high accuracy without high costs and workflow disruptions. In addition, the system eliminates the need for surgeons to refer to separate monitors displaying the preoperative plan, a limitation inherent in traditional methods that rely on fluoroscopy and computer-assisted navigation systems. 

Modular jigs offer a cost-effective alternative to patient-specific instruments by significantly reducing both production time and expenses. Unlike conventional 3D-printed jigs, which can take weeks to fabricate and are disposed after single use, modular jigs utilize standardized components that can be quickly assembled and reused for multiple cases after sterilization. This flexibility not only shortens the lead time but also minimizes the need for customized manufacturing, which can be prohibitively expensive, for example, in the United States. For instance, a single case using a 3D-printed jig might cost around \$20,000 (Onkos Surgical, June 2018), as seen in some distal femur surgeries. In contrast, modular jigs eliminate the need for such high expenditures by allowing surgeons to adapt the jig to various surgical scenarios without the need for new, patient-specific designs each time. This adaptability not only reduces the immediate costs associated with each surgery but also lowers long-term expenses by enabling the reuse of components, making modular jigs a sustainable and economically viable option in surgical practice.

The results of this study should be considered with several limitations. Although promising, this technology is still in the proof-of-concept stage with experiments conducted using simulated pelvic bone tumor resections on synthetic bones. Further adaptations are needed for sterile compliance in operating rooms. While sawbone models were used for initial testing, a cadaver study will confirm the system's effectiveness in more realistic conditions. Although this system has been tested on a cadaver of the right lower extremity, including the femur and half tibia~\cite{he2022light-jigs}, a clinical study using a cadaver pelvis is still required.

The system's structured-light scanner captures extensive data points to enhance registration accuracy compared to traditional methods. However, it requires a clear line-of-sight, which can be challenging during surgery. To address this, we developed an impression molding technique for accurate registration even with obstructions in the field of view~\cite{he2022novelVPS}.

Although the freehand method demonstrated relatively precise resections compared to other studies~\cite{cartiaux2008surgical}, this precision was largely facilitated by the use of sawbones without overlying soft tissues. This setup allowed the surgeon to make precise measurements and view the pelvis from angles that are not typically possible during actual surgery. In addition, the surgeon was able to perform the freehand surgery with meticulous care, taking extra time to accurately trace the osteotomy lines. Such precision and time allocation are often impractical in actual surgeries, where time constraints and the presence of soft tissues can significantly impact the accuracy of freehand resections.

During pelvic bone tumor surgeries, achieving a complete resection of a tumor involving the acetabulum area traditionally requires three osteotomies. The proposed modular jig design effectively assists the surgeon in performing supra-acetabular and infra-acetabular osteotomies, offering a significant advantage over the traditional freehand method by preserving more healthy bone. However, the osteotomy of the superior pubic ramus presents a challenge due to the limited bony area available for securing a standard jig. In these situations, the vision-guided system can project the osteotomy line directly onto the bone, providing the surgeon with precise guidance for resection. This capability enhances the surgeon's ability to execute accurate resections even in anatomically constrained areas, thereby improving surgical outcomes.

In summary, the vision-guided and modular jig system presents a promising, cost-effective alternative to traditional methods. This innovative approach has the potential to enhance pelvic bone tumor resection by improving surgical precision, reducing complications, and enhancing patient outcomes while addressing the limitations of existing technologies. Ongoing research will focus on expanding its capabilities and conducting clinical trials to confirm its efficacy in diverse surgical scenarios. Uncertainty analysis of the system will be crucial in identifying potential sources of error and variability, allowing for further refinement and optimization of the technology~\cite{he2023uncertainty, he2024uncertainty}. This analysis will help ensure that the system consistently delivers accurate and reliable results, ultimately contributing to its successful integration into clinical practice.


\section{Conclusion}
This study demonstrates that the integration of a real-time vision-guided surgical system with modular cutting jigs offers a promising solution for enhancing the precision of pelvic tumor resections. The system successfully addresses key limitations of current approaches by providing accurate guidance without the drawbacks of high costs, radiation exposure, workflow disruptions, or long production times that are typically associated with navigation systems and patient-specific instruments. The system achieved substantial improvements across all measured parameters, with mean distance deviation reduced by 51.2\%, roll and pitch angle deviation decreased by 72.6\% and 70.2\%, respectively.
The modular nature of the cutting jigs, combined with real-time optical guidance, presents a scalable and practical solution that can be readily implemented in clinical settings.

While experiment results are promising, further validation through cadaveric studies and clinical trials is necessary to confirm the system's effectiveness in actual surgical conditions. Future developments will focus on expanding the system's capabilities to accommodate diverse resection geometries, ensuring seamless integration into sterile operating room environments, and validating its performance across a broader range of pelvic bone tumor cases. This technology shows potential to improve surgical outcomes while making precise bone tumor resection techniques more accessible to a wider range of medical facilities.

\section*{Acknowledgment}
We would like to thank Charles K. Mazzarese, MPS from the Department of Radiology at the Renaissance School of Medicine, Stony Brook University, for his assistance with this project.

\section*{Conflict of Interest}
Imin Kao and Fazel A. Khan have a patent US patent application No. 16/854,804 licensed to NaviSect.
Fazel A. Khan is part owner of the NaviSect, INC.

\section*{Author Contribution}
\textbf{Vahid Danesh} designed the study, drafted the manuscript, and contributed substantially to data acquisition, analysis, and interpretation. \textbf{Paul Arauz} was responsible for study design, data acquisition, interpretation, and critical review of the manuscript. \textbf{Maede Boroji},\textbf{Andrew Zhu}, and \textbf{Mia Cottone} contributed to data acquisition, interpretation, and critical review of the manuscript. \textbf{Elaine Gould} contributed to data acquisition. \textbf{Fazel A. Khan} and \textbf{Imin Kao} were involved in research design, data interpretation, and critical review of the manuscript. All authors have read and approved the final submitted manuscript.


\bibliography{refs}

\begin{thebibliography}{10}
\providecommand \doibase [0]{http://dx.doi.org/}%

\bibitem{ozaki2003osteosarcoma}
Ozaki T, Flege S, Kevric M, et al. Osteosarcoma of the pelvis: experience of the Cooperative Osteosarcoma Study Group. {\it Journal of clinical oncology.} 2003\string;21(2)\string:334--341.
\newblock \href {\doibase 10.1200/JCO.2003.01.142} {doi: 10.1200/JCO.2003.01.142}

\bibitem{fuchs2009osteosarcoma}
Fuchs B, Hoekzema N, Larson DR, Inwards CY, Sim FH. Osteosarcoma of the pelvis: outcome analysis of surgical treatment. {\it Clinical orthopaedics and related research.} 2009\string;467\string:510--518.
\newblock \href {\doibase 10.1007/s11999-008-0495-x} {doi: 10.1007/s11999-008-0495-x}

\bibitem{hoffer2000}
Hoffer FA, Nikanorov AY, Reddick WE, et al. Accuracy of MR imaging for detecting epiphyseal extension of osteosarcoma. {\it Pediatric radiology.} 2000\string;30\string:289-298.
\newblock \href {\doibase 10.1007/s002470050743} {doi: 10.1007/s002470050743}

\bibitem{saifuddin2002}
Saifuddin A. The accuracy of imaging in the local staging of appendicular osteosarcoma. {\it Skeletal radiology.} 2002\string;31\string:191-201.
\newblock \href {\doibase 10.1007/s00256-001-0471-y} {doi: 10.1007/s00256-001-0471-y}

\bibitem{khan2013}
Khan F, Pearle A, Lightcap C, Boland PJ, Healey JH. Haptic robot-assisted surgery improves accuracy of wide resection of bone tumors: a pilot study. {\it Clinical Orthopaedics and Related Research®.} 2013\string;471\string:851-859.
\newblock \href {\doibase 10.1007/s11999-012-2529-7} {doi: 10.1007/s11999-012-2529-7}

\bibitem{springfield1984}
Springfield D, Enneking W, Neff J, Makley J. Principles of tumor management. {\it Instructional course lectures.} 1984\string;33\string:1-25.

\bibitem{springfield1988}
Springfield DS, Schmidt R, Graham-Pole J, Marcus~Jr R, Spanier S, Enneking W. Surgical treatment for osteosarcoma. {\it JBJS.} 1988\string;70(8)\string:1124-1130.

\bibitem{cartiaux2008surgical}
Cartiaux O, Docquier PL, Paul L, et al. Surgical inaccuracy of tumor resection and reconstruction within the pelvis: an experimental study. {\it Acta orthopaedica.} 2008\string;79(5)\string:695--702.
\newblock \href {\doibase 10.1080/17453670810016731} {doi: 10.1080/17453670810016731}

\bibitem{cartiaux2010computer}
Cartiaux O, Paul L, Docquier PL, Raucent B, Dombre E, Banse X. Computer-assisted and robot-assisted technologies to improve bone-cutting accuracy when integrated with a freehand process using an oscillating saw. {\it JBJS.} 2010\string;92(11)\string:2076--2082.
\newblock \href {\doibase 10.2106/JBJS.I.00457} {doi: 10.2106/JBJS.I.00457}

\bibitem{bacci2005}
Bacci G, Briccoli A, Longhi A, et al. Treatment and outcome of recurrent osteosarcoma: experience at Rizzoli in 235 patients initially treated with neoadjuvant chemotherapy. {\it Acta Oncologica.} 2005\string;44(7)\string:748-755.
\newblock \href {\doibase 10.1080/02841860500327503} {doi: 10.1080/02841860500327503}

\bibitem{nathan2006}
Nathan SS, Gorlick R, Bukata S, et al. Treatment algorithm for locally recurrent osteosarcoma based on local disease‐free interval and the presence of lung metastasis. {\it Cancer.} 2006\string;107(7)\string:1607-1616.
\newblock \href {\doibase 10.1002/cncr.22197} {doi: 10.1002/cncr.22197}

\bibitem{avedian2010}
Avedian RS, Haydon RC, Peabody TD. Multiplanar osteotomy with limited wide margins: a tissue preserving surgical technique for high-grade bone sarcomas. {\it Clinical Orthopaedics and Related Research®.} 2010\string;468(10)\string:2754-2764.
\newblock \href {\doibase 10.1007/s11999-010-1362-0} {doi: 10.1007/s11999-010-1362-0}

\bibitem{cartiaux2013}
Cartiaux O, Banse X, Paul L, Francq BG, Aubin CE, Docquier PL. Computer-assisted planning and navigation improves cutting accuracy during simulated bone tumor surgery of the pelvis. {\it Computer Aided Surgery.} 2013\string;18(1-2)\string:19-26.
\newblock \href {\doibase 10.3109/10929088.2012.744096} {doi: 10.3109/10929088.2012.744096}

\bibitem{cartiaux2011}
Cartiaux O, Paul L, Docquier PL, Banse X. Computer-and robot-assisted resection and reconstruction of pelvic bone tumours—a review. {\it European Musculoskeletal Review.} 2011\string;6(2)\string:125-130.
\newblock \href {\doibase 10.1097/BCO.0b013e318221b1a3} {doi: 10.1097/BCO.0b013e318221b1a3}

\bibitem{abraham2011}
Abraham JA. Recent advances in navigation-assisted musculoskeletal tumor resection. {\it Current Orthopaedic Practice.} 2011\string;22(4)\string:297-302.

\bibitem{fehlberg2009}
Fehlberg S, Eulenstein S, Lange T, Andreou D, Tunn PU. Computer-assisted pelvic tumor resection: fields of application, limits, and perspectives. {\it Treatment of Bone and Soft Tissue Sarcomas.} 2009\string:169-182.
\newblock \href {\doibase 10.1007/978-3-540-77960-5\_11} {doi: 10.1007/978-3-540-77960-5\_11}

\bibitem{garcia2021}
García-Sevilla M, Mediavilla-Santos L, Moreta-Martinez R, et al. Combining Surgical Navigation and 3D Printing for Less Invasive Pelvic Tumor Resections. {\it IEEE Access.} 2021\string;9\string:133541-133551.
\newblock \href {\doibase 10.1109/ACCESS.2021.3115984} {doi: 10.1109/ACCESS.2021.3115984}

\bibitem{biscaccianti2022}
Biscaccianti V, Fragnaud H, Hascoët JY, Crenn V, Vidal L. Digital chain for pelvic tumor resection with 3D-printed surgical cutting guides. {\it Frontiers in Bioengineering and Biotechnology.} 2022\string;10\string:991676.
\newblock \href {\doibase 10.3389/fbioe.2022.991676} {doi: 10.3389/fbioe.2022.991676}

\bibitem{cartiaux2014}
Cartiaux O, Paul L, Francq BG, Banse X, Docquier PL. Improved accuracy with 3D planning and patient-specific instruments during simulated pelvic bone tumor surgery. {\it Annals of biomedical engineering.} 2014\string;42\string:205-213.
\newblock \href {\doibase 10.1007/s10439-013-0890-7} {doi: 10.1007/s10439-013-0890-7}

\bibitem{fragnaud2022}
Fragnaud H, Biscaccianti V, Hascoët JY, et al. How does customized cutting guide design affect accuracy and ergonomics in pelvic tumor resection? A study in cadavers. {\it Clinical Orthopaedics and Related Research®.} 2022\string:10.1097.
\newblock \href {\doibase https://doi.org/10.1097/corr.0000000000003000} {doi: https://doi.org/10.1097/corr.0000000000003000}

\bibitem{jentzsch2016}
Jentzsch T, Vlachopoulos L, Fürnstahl P, Müller DA, Fuchs B. Tumor resection at the pelvis using three-dimensional planning and patient-specific instruments: a case series. {\it World journal of surgical oncology.} 2016\string;14\string:1-12.
\newblock \href {\doibase 10.1186/s12957-016-1006-2} {doi: 10.1186/s12957-016-1006-2}

\bibitem{sallent2017}
Sallent A, Vicente M, Reverté M, et al. How 3D patient-specific instruments improve accuracy of pelvic bone tumour resection in a cadaveric study. {\it Bone \& joint research.} 2017\string;6(10)\string:577-583.
\newblock \href {\doibase 10.1302/2046-3758.610.BJR-2017-0094.R1} {doi: 10.1302/2046-3758.610.BJR-2017-0094.R1}

\bibitem{evrard2019}
Evrard R, Schubert T, Paul L, Docquier PL. Resection margins obtained with patient-specific instruments for resecting primary pelvic bone sarcomas: A case-control study. {\it Orthopaedics \& Traumatology: Surgery \& Research.} 2019\string;105(4)\string:781-787.
\newblock \href {\doibase 10.1016/j.otsr.2018.12.016} {doi: 10.1016/j.otsr.2018.12.016}

\bibitem{mavrogenis2012}
Mavrogenis AF, Soultanis K, Patapis P, et al. Pelvic resections. {\it Orthopedics.} 2012\string;35(2)\string:e232-e243.
\newblock \href {\doibase 10.3928/01477447-20120123-40} {doi: 10.3928/01477447-20120123-40}

\bibitem{dorling2023cost}
Dorling IM, Geenen L, Heymans MJ, Most J, Boonen B, Schotanus MG. Cost-effectiveness of patient specific vs conventional instrumentation for total knee arthroplasty: A systematic review and meta-analysis. {\it World Journal of Orthopedics.} 2023\string;14(6)\string:458.

\bibitem{he2022light-jigs}
He G, Dai AZ, Mustahsan VM, et al. A novel method of light projection and modular jigs to improve accuracy in bone sarcoma resection. {\it Journal of Orthopaedic Research®.} 2022\string;40(11)\string:2522-2536.
\newblock \href {\doibase 10.1002/jor.25300} {doi: 10.1002/jor.25300}

\bibitem{he2022light3d}
He G, Dai AZ, Mustahsan VM, Blum CL, Kao I, Khan FA. A novel 3d light assisted drawing (3d-lad) method to aid intraoperative reproduction of osteotomy lines surrounding a bone tumor during wide resection: An experimental study. {\it Orthopedic Research and Reviews.} 2022\string:101-109.
\newblock \href {\doibase 10.2147/ORR.S349240} {doi: 10.2147/ORR.S349240}

\bibitem{he2022novelVPS}
He G, Ricca JM, Dai AZ, et al. A novel bone registration method using impression molding and structured-light 3D scanning technology. {\it Journal of Orthopaedic Research{\textregistered}.} 2022\string;40(10)\string:2340--2349.
\newblock \href {\doibase 10.1002/jor.25275} {doi: 10.1002/jor.25275}

\bibitem{he2021registration}
He G, Mustahsan VM, Bielski MR, Kao I, Khan FA. Report on a novel bone registration method: a rapid, accurate, and radiation-free technique for computer-and robotic-assisted orthopedic surgeries. {\it Journal of Orthopaedics.} 2021\string;23\string:227-232.

\bibitem{Bickels2001}
Bickels J, Malawer M. {\it Pelvic Resections (Internal Hemipelvectomies)}\string:405--414; Dordrecht: Springer Netherlands .
\newblock 2001

\bibitem{Malawer2001}
Malawer M. {\it Periacetabular Resections}\string:425--438; Dordrecht: Springer Netherlands .
\newblock 2001

\bibitem{he2023uncertainty}
He G, Fakhari A, Khan F, Kao I. Experimental and Computational Study of Error and Uncertainty in Real-Time Camera-Based Tracking of a Two-Dimensional Marker for Orthopedic Surgical Navigation. {\it Journal of Verification, Validation and Uncertainty Quantification.} 2023\string;8(2)\string:021001.
\newblock \href {\doibase 10.1115/1.4062137} {doi: 10.1115/1.4062137}

\bibitem{he2024uncertainty}
He G, Fakhari A, Khan F, Kao I. Propagation of Error and Uncertainty in a Computer-Assisted Orthopedic Surgical System. {\it IEEE Transactions on Instrumentation and Measurement.} 2024\string;73\string:1-15.
\newblock \href {\doibase 10.1109/TIM.2024.3378299} {doi: 10.1109/TIM.2024.3378299}

\end{thebibliography}

\appendix
\bmsection{Design and Assembly of Standard Modular Jig}
\label{app:modular_jig}
3D modeling software, Geomagic Design X (3D Systems Corp, Rock Hill, SC, USA), and Fusion 360 (Autodesk Inc, San Francisco, CA, USA), were utilized to evaluate pelvic geometry and plan potential resection approaches for Type II pelvic bone tumors.

The collaborative engineering and medical team designed and fabricated multiple standardized modular jig components to allow precise tumor resection through various assembly configurations. These standardized components can be assembled quickly and easily during surgery, allowing surgeons to adapt the modular jig to achieve the necessary resection angles and dimensions. This ensures the jig conforms both to the patient’s anatomy and the preoperative resection plan. In this study, the standardized jig components were 3D printed using a Formlabs Form 2 3D printer (Formlabs, Inc., Somerville, MA, USA). Eventually, these components will be machined out of medical-grade metals which can be reused.

\begin{figure}
    \centering
    \includegraphics[width=0.95\linewidth]{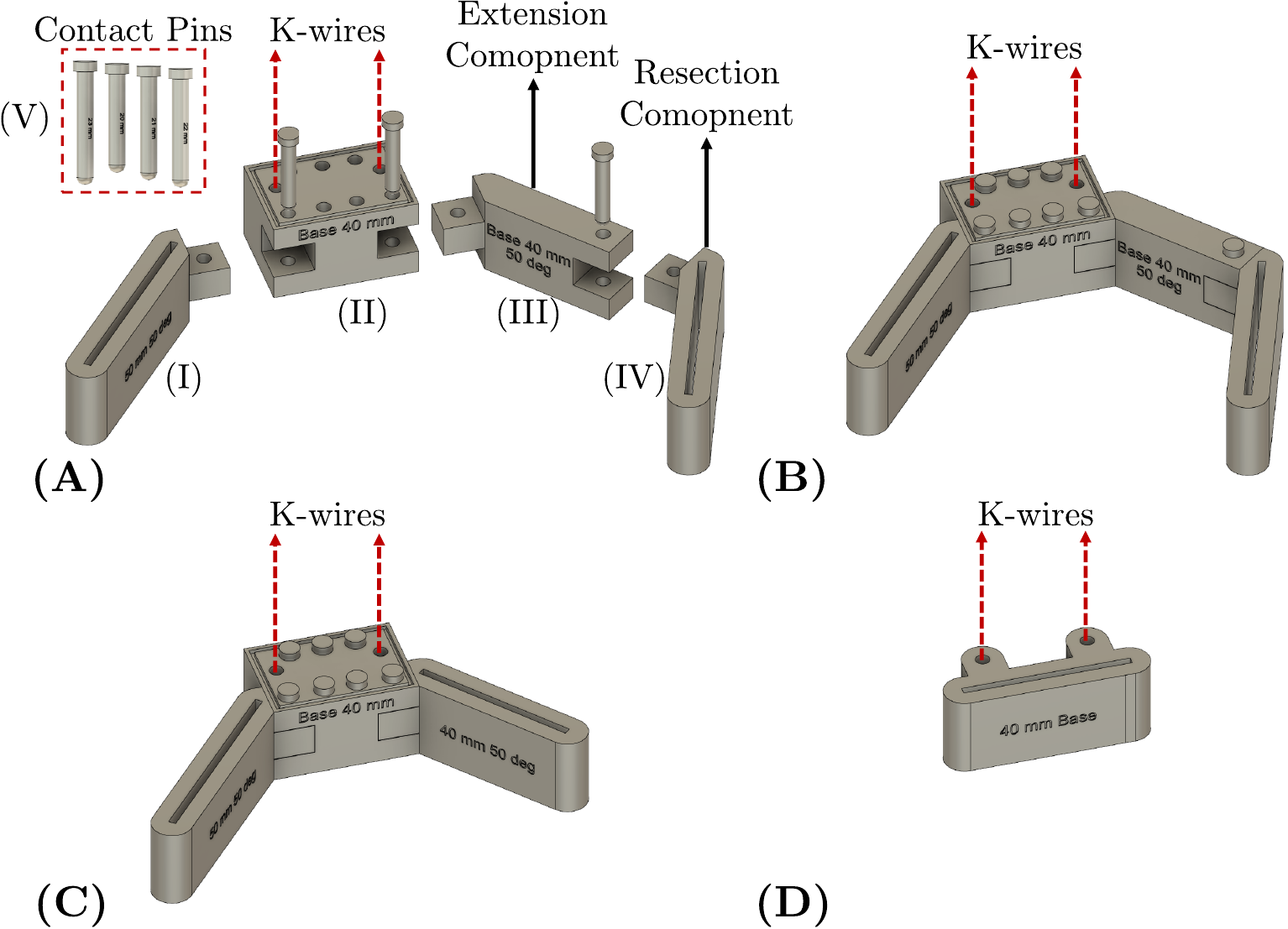}
    \caption{Assembly and configuration of a modular jig. 
    (A) Displays the main components of the modular jig: (I) first resection component, (II) base component, (III) extension component, (IV) second resection component, and (V) contact pins. 
    (B) The assembled jig of the components in (A);
    (C) Illustrates the jig after replacing the extension component by the third resection component; 
    (D) The fourth resection component, completing the setup for precise resection.}
    \label{AppFig:Jig}
\end{figure}

Discrete stock sizes and angles are available to accommodate a wide range of tumor sizes. In addition, snap-fit pins with a standard diameter but varying lengths are used to optimize jig positioning, allowing the modular cutting jig to establish secure contact with multiple bone points, enhancing accuracy and stability. For the Type II cases, the modular jig design featured a 35$\times$24$\times$20 mm$^3$ rectangular prism base component, shown in Fig~\ref{AppFig:Jig}(A)-(II). This base included two holes for guiding K-wires for fixation, multiple holes for snap-fit pins to ensure stable contact with the bone, an engraved rectangular pattern for alignment with the light projected pattern, and two mating slots for attaching resection and/or extension components. The design also included four resection components, each with a 10mm width, 20mm height, and a 1.8mm thick slot aligned with the cutting plane. Two of these resection components attach directly to the base, while an extension component connects the base to another resection component. The final resection component has two holes that align with the K-wire holes in the base but remains unconnected to other components.

The modular jig assembly, placement, and resection sequence consist of the following steps (Fig~\ref{AppFig:Jig}):
\begin{enumerate}
    \item Connect the base to the first resection component and an extension component, then attach the second resection component through the extension. Contact pins are preoperatively selected and inserted into the base to optimize jig positioning on the bone, as shown in Fig~\ref{AppFig:Jig}(A) and Fig~\ref{AppFig:Jig}(B).
    \item Position the jig on the bone to align the engraved pattern on the jig base with projected pattern. Secure the jig with two parallel K-wires, then perform the first two cuts.
    
    \item Replace the extension with the third resection component and perform the third cut, as shown in Fig~\ref{AppFig:Jig}(C).
    
    \item Replace the base with the final resection jig for the final cut, as shown in Fig~\ref{AppFig:Jig}(D).
\end{enumerate}

This modular approach ensures precise and customizable resections based on individual anatomy and tumor characteristics.

\end{document}